\documentclass[preprint,12pt]{elsarticle}

\usepackage{amssymb,amsfonts,amsmath}
\usepackage{graphicx}
\usepackage{dcolumn}
\usepackage{color}

\journal{Physica A}

\begin{document}

\begin{frontmatter}

\title{On alternative formulations of the small-world metric in complex networks}

\author{Massimiliano Zanin}

\address{Innaxis Foundation \& Research Institute,
Jos\'e Ortega y Gasset 20, 28006, Madrid, Spain}
\address{Faculdade de Ci\^encias e Tecnologia, Departamento de Engenharia Electrot\'ecnica,
Universidade Nova de Lisboa, Lisboa, Portugal}

\begin{abstract}
Small-world networks, {\it i.e.} networks displaying both a high clustering coefficient and a small characteristic path length, are obliquitous in nature. Since their identification, the ``small-worldness'' metric, as proposed by Humphries and Gurney, has frequently been used to detect such structural property in real-world complex networks, to a large extent in the study of brain dynamics. Here I discuss several of its drawbacks, including its lack of definition in disconnected networks and the impossibility of assessing a statistical significance; and present different alternative formulations to overcome these difficulties, validated through the phenospaces representing a set of $48$ real networks.
\end{abstract}

\begin{keyword}
Complex networks \sep small-worldness \sep efficiency
\end{keyword}

\end{frontmatter}

\section{Introduction}

Small-worldness \cite{watts1998collective} is a complex network property that has received a huge amount of interest in the last decade. Artificial networks were initially created around two different paradigms: random graphs, in which the existence of a link is the result of a random process \cite{bollobas1998random}; and regular ones, whose nodes have the same number of connections. It was nevertheless soon discovered that many real technological, biological and social networks fall in the middle: they show high local clustering ({\it e.g.} triangles), like regular networks, but also short path lengths between elements, characteristic of random graphs.

With the aim of providing an objective measure of the small-world nature of real networks, Humphries and Gurney \cite{Humphries:2008kn} proposed in 2008 a metric based on the celebrated Watts-Strogatz (WS) model. The {\it small-worldness} structural measure $S$ was defined as the ratio between the clustering coefficient and the characteristic paths length, normalised according to the values expected in random equivalent networks:

\begin{equation}
	S = \frac{C}{C_{rand}} \frac{L_{rand}}{L}.
	\label{eq:S}
\end{equation}

Here, $C_{rand}$ and $L_{rand}$ respectively represent the clustering coefficient and the characteristic paths length observed in random equivalent networks, {\it i.e.} random networks with the same number of nodes and links than the one under study. As a network is said to be small-world when $L \approx L_{rand}$ and $C \gg  C_{rand}$, $S > 1$ indicates the presence of such property.

Since its introduction, the small-world metric as defined in Eq.~\ref{eq:S} has been applied to the study of a large number of real systems, with a special attention devoted to the human brain, both in normal and pathological conditions \cite{bullmore2009complex, rubinov2010complex}.
The reasons for such interest become clear when one highlights the fact that the small-worldness synthesises two important aspects of networks (and of brain) dynamics: local interconnectivity, through $C$, as the creation of group of nodes strongly and redundantly connected between them; and global integration, through $L$, representing the movement of information across large distances.
Such balance between short- and long-range connectivities is altered in the Alzheimer's disease, both in patients \cite{supekar2008network, sanz2010loss} and in control subjects carrying genetic variations used as biomarkers \cite{brown2011brain}; similar results were also obtained for individuals suffering from Mild Cognitive Impairment \cite{yao2010abnormal, wang2013disrupted}, the prodromal stage of Alzheimer's. Beyond biology, small-worldness has been applied to the analysis of terrorists social networks \cite{everton2009network}, of audio clip sharing communities \cite{roma2012small}, up to as a criteria for organising datacenters \cite{shin2011small}, among others. 

In spite of its popularity, such metric presents several drawbacks. First, it is defined only for connected networks; when disconnected components are present, a common situation in real systems, $L$ diverges to infinity. Second, the normalisation with respect to random networks does not yield information about the significance of the obtained value. In this contribution, I address these two problems by presenting two alternative formulations of the small-world metric. They are respectively based on the concepts of {\it efficiency} (Section \ref{sec:efficiency}), which, while conveying the same information as $L$, is defined even in the case of disconnected networks; and of {\it ZScore} (Section \ref{sec:zscore}), enabling a better estimation of the statistical significance of results. Both metrics are then tested against a set of $48$ real networks, covering social, biological and technological systems \cite{killworth1976informant, hummon1989connectivity, wasserman1994social, batagelj2006pajek, bu2003topological, lusseau2003emergent, bascompte2005simple, opsahl2010node}. Section \ref{sec:concl} finally draws some conclusions, and recommendations on the use of the small-worldness for the analysis of real networks.

\section{Efficiency {\it vs.} Characteristic paths length}
\label{sec:efficiency}

The first problem in the application of the small-world metric appears when the network under analysis is not connected, {\it i.e.} when a path cannot be constructed between some of its nodes. For those pairs, the characteristic paths length $L$ diverges, and thus $S \rightarrow 0$. Even if the original graph is composed of a single component, the random networks used for normalisation may be disconnected, especially when the probability of links appearance $p$ is below the threshold $\frac{\ln n}{n}$, $n$ being the number of nodes \cite{bollobas1998random}.

It is worth noticing that this situation frequently appears in the study of real systems, and especially in the study of brain dynamics. For instance, the functional network \cite{bullmore2009complex} representing a given cognitive task may not connect all brain regions, as some of them may not be involved in the computation. Two solutions can then be adopted. First, consider only the giant component of the network, {\it i.e.} the largest group of nodes forming a connected sub-graph; resulting networks may nevertheless have different number of nodes, and represent different parts of the brain, making difficult any comparison between subjects and tasks. Second, networks can be created by applying dynamical thresholds, in order to ensure the connectivity of the network, at the price of possibly including links without biological value \cite{papo2014functional}.

Here I propose a different approach, which stems from the use of a distance metric that is well defined even for disconnected networks. Such metric, called {\it Efficiency} \cite{latora2001efficient, latora2002boston}, is defined as:

\begin{equation}
	E(\mathcal G) = \frac{1}{n (n-1)} \sum _{i \neq j \in \mathcal G} \frac{1}{d_ij}.
\end{equation}

The efficiency is thus the inverse of the harmonic mean of all shortest paths lengths $d_{ij}$, being $i$ and $j$ nodes of the graph $\mathcal G$, normalised in order to obtain $0 \leq E \leq 1$. Notice that when the graph is fully disconnected, $d_{ij} \rightarrow \infty$ and thus $E \rightarrow 0$. Being the efficiency inversely proportional to the shortest distance between nodes, {\it i.e.} $E \approx 1 / L$, the small-world metric can be redefined as:

\begin{equation}
	S^E = \frac{C}{C_{rand}} \frac{E}{E_{rand}},
\end{equation}

which has the advantage of being defined independently of the connectedness of the network.

\begin{figure}[!tb]
\begin{center}
\includegraphics[width=0.99\textwidth]{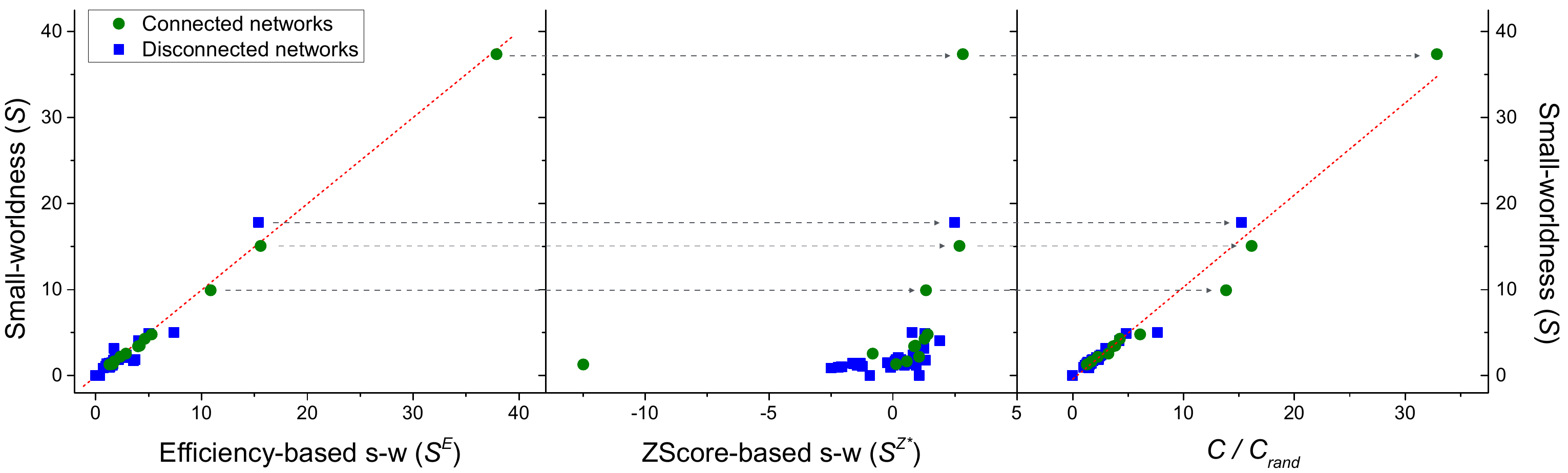}
\caption{Comparison between the small-worldness $S$ and its alternative formulations, through the phenospace of $48$ real networks. (Left) $S$ {\it vs.} $S^E$, with the red dashed line representing the identity $S = S^E$. (Center) $S$ {\it vs.} $S^{Z*}$. (Right) $S$ {\it vs.} $C_{rand}$. In all cases, green circles and blue squares respectively represent connected and disconnected networks; for the latter, $S$ has been calculated over their giant component. The Y axis is common to all three graphs, allowing to follow the same network through them, as symbolised by the horizontal dashed arrows.
}\label{fig:01}
\end{center}
\end{figure}

Fig. \ref{fig:01} Left compares $S$ and $S^E$ through the phenospace created by the $48$ real networks here considered, where each point in the plane represents a network, and its coordinates are given by the values of both metrics. In the case of disconnected networks (blue squares throughout Fig. \ref{fig:01}), $S$ has been calculated only on their giant connected component. It can be appreciated that, for connected networks (green circles), $S^E$ very well approximates $S$ (notice the red dashed line corresponding to $S = S^E$). On the other hand, disconnected networks deviate from the identity relation, both above and below. In the latter case thus $S$ is biased by the removal of loosely connected nodes, which decreases both $L$ (thus increasing $S$) and $C$ (decreasing $S$). $S^E$ maintains the original meaning of $S$, while providing a better assessment of the structure of disconnected networks.

\section{ZScore {\it vs.} expected values}
\label{sec:zscore}

The second problem here addressed is the definition of a normalisation procedure that can yield information about the statistical significance of results. According to Eq. \ref{eq:S}, this is usually performed by normalising $C$ and $L$ according to their expected values $C_{rand}$ and $L_{rand}$, as observed in an ensemble of random networks with the same number of nodes and links. While $C / C_{rand}$ and $L / L_{rand}$ represent how far are the obtained values from what expected, they do not provide information about the statistical relevance of such values. Suppose two networks $\mathcal G_1$ and $\mathcal G_2$, such that $C(\mathcal G_1) = 0.05$, $C_{rand}(\mathcal G_1) = 0.025$, $C(\mathcal G_2) = 1$ and $C_{rand}(\mathcal G_2) = 0.5$. In both cases,

\begin{equation}
	\frac{CC(\mathcal G_1)}{CC_{rand}(\mathcal G_1)} = \frac{CC(\mathcal G_2)}{CC_{rand}(\mathcal G_2)} = 2.0.
\end{equation}

An attentive eye would nevertheless observe that the clustering coefficient of $\mathcal G_2$ is more unusual, as a perfect clusterisation $C(\mathcal G_2) = 1$ is hardly expected in random networks.

In order to highlight the difference between both situations, and thus to assess the statistical significance of $S$, I here propose the use of the {\it ZScore}, a standard method for calculating the $p$-value of a measurement given a Gaussian reference distribution:

\begin{equation}
	ZScore(M) = \frac{M - \langle {M_{rand}} \rangle }{\sigma(M_{rand})}.
\end{equation}

$M$ represents the metric under analysis, $M_{rand}$ a set of values obtained in random equivalent networks, and $\langle \cdot \rangle$ and $\sigma(\cdot)$ respectively their average and standard deviation. Large positive and negative values of the ZScore (respectively above $2$ and below $-2$) represent statistically significant high and low observed values. The small-world metric can then be reformulated as follows:

\begin{equation}
	S^Z = ZScore(CC) - ZScore(L);
\end{equation}

or, including the efficiency in the definition, as:

\begin{equation}
	S^{ZE} = ZScore(CC) + ZScore(E).
\end{equation}

Fig. \ref{fig:01} Center depicts the relation between $S$ and $S^Z$. As ZScore values can assume extreme values, it is here represented as its logarithm, {\it i.e.} $S^{Z*} = sign ( S^Z ) \cdot \log _{10} | S^Z |$. It can be noticed that similar values of $S$, {\it e.g.} between $0.0$ and $1.0$, can have very different statistical significances ($S^{Z*}$ between $0$ and $-12$, {\it i.e.} $S^Z$ between $0$ and $-10^{12}$, the latter corresponding to extremely small $p$-values).

\section{Conclusions and discussion}
\label{sec:concl}

\begin{figure}[!tb]
\begin{center}
\includegraphics[width=0.99\textwidth]{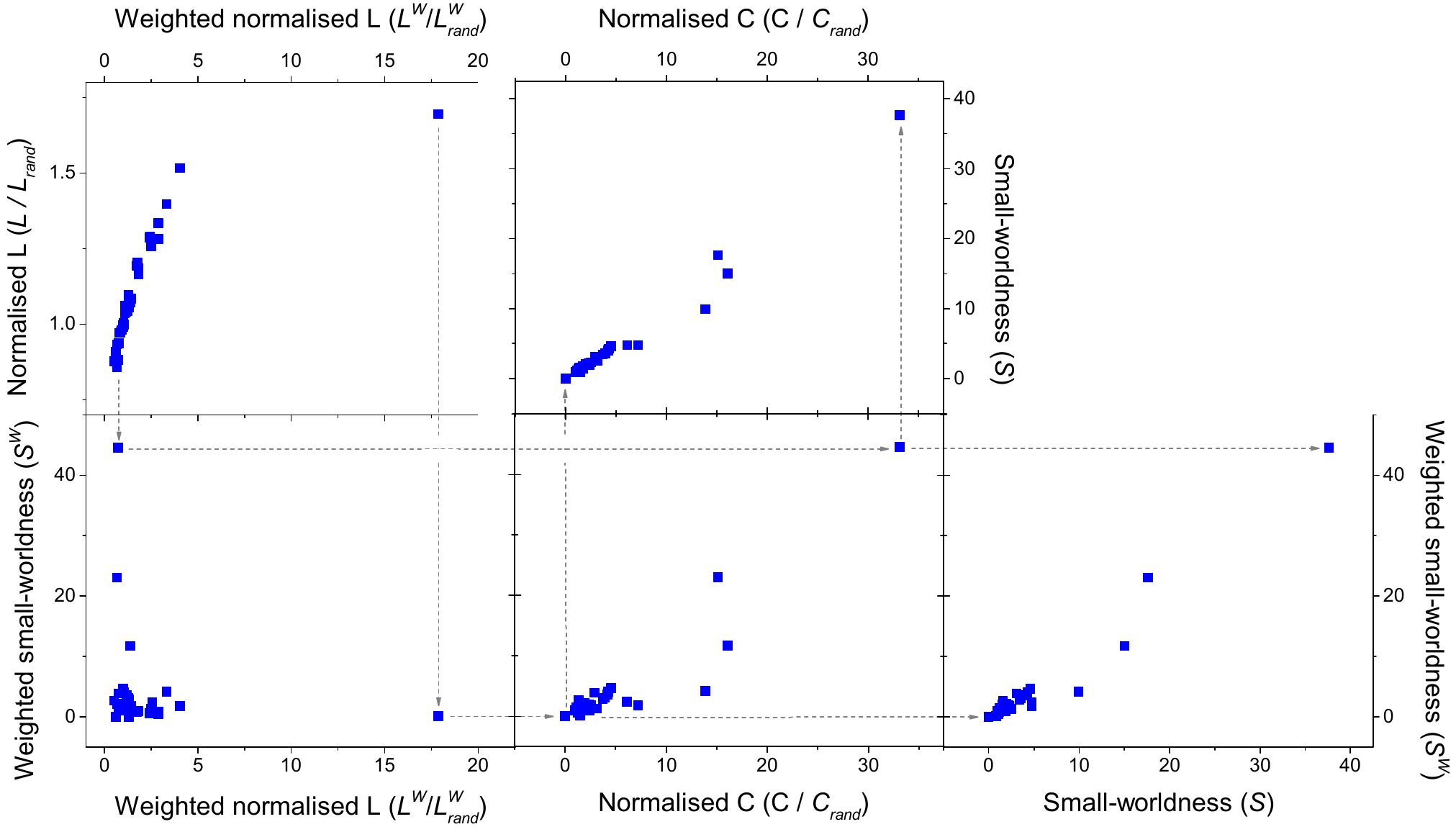}
\caption{Small-worldness $S$ and its weighted version $S^W$. Through the different panels composing this Figure, it is possible to compare pairs of metrics, including $S$, $L$, $C$, and their weighted version $S^W$ and $L^W$. When the X and Y axes are common to two panels, it is possible to follow the same network through them, as symbolised by the dashed grey arrows.
}\label{fig:02}
\end{center}
\end{figure}

This contribution discusses the problem of numerically evaluating the small-world property of a complex network, {\it i.e.} identifying structures that are characterised both by a high clustering coefficient and a small shortest path distance between nodes. The metric originally proposed by Humphries and Gurney \cite{Humphries:2008kn} presents two main drawbacks, namely the inability to handle disconnected networks, and the little information provided about the statistical significance of results. Here I show how these two problems can be solved by respectively including the efficiency of the network, as a proxy of the inverse of the geodesic distance between nodes; and by using a ZScore instead of a simple normalisation, which allows assessing the $p$-value of the observations.

Beyond the problem of correctly quantifying the presence of a small-world structure, a researcher analysing real networks should also be careful about understanding the origin of that property. Independently on the specific metric used, the calculation of $S$ (as well as of $S^E$ and $S^Z$) implies a reduction of the information available, from a two-dimension space (given by $C$ and $L$, or $C$ and $E$) to an unidimensional one: information is thus always lost in the process. One remaining task is thus to understand from what network property the small-worldness arises, {\it i.e.} from a higher than expected clustering coefficient, or from a smaller than expected long-range connectivity. In order to shed light on this issue, Fig. \ref{fig:01} Right depicts the relation between $S$ and $C_{rand}$, as observed in the $48$ real networks here considered. The red line represents the best linear fit between them ($S = 1.07 \cdot C - 0.42$, $R^2 = 0.972$): the small-world property is thus largely explained by the clustering coefficient, making the former metric mostly redundant in the understanding of those systems. This problem is especially relevant in small networks, as in the representations created from EEG and MEG recordings of brain activity \cite{bullmore2009complex}, in which $L$ is largely constrained.

I here suggest to solve this latter problem by including, within the definition of the small-worldness, a weighted connectivity metric that strongly penalises the presence of pairs of nodes whose distance is greater than the one expected in a small-world network, {\it i.e.} $d > \ln n$. This requires, first, to normalise the distance between pairs of nodes as follows:

\begin{equation}
	d_{i,j}^{\ln} = d_{i,j} / \ln n,
\end{equation}

being $n$ the number of nodes in the network. $L$ can then be redefined as:

\begin{equation}
	L^W = \frac{1}{n(n-1)} \sum _{i, j \neq i} ( d_{i,j}^{\ln} ) ^w,
\end{equation}

$w > 1$ being a parameters used to penalise long paths, {\it i.e.} those longer than $\ln n$. The small-worldness metric can then be updated accordingly, by introducing the weighted measures $L^W$ and $L^W_{rand}$ inside Eq. \ref{eq:S}. Fig. \ref{fig:02} depicts, throughout its panels, the phenospaces created by pairs of metrics, including both standard ($C$, $L$ and $S$) and weighted ones ($L^W$ and $S^W$), for $w = 3$. Of special relevance are the two central panels, in which $S$ (top) and $S^W$ (bottom) are compared with the normalised clustering coefficient; it can be appreciated that the weighted small-worldness introduces more variability, $C$ being no longer enough to explain the network structure.

In summary, the researcher dealing with small-world complex networks, and willing to quantifying such topological structure, should choose the metric accordingly to the property of the network, and specifically its connectedness and its size. MATLAB\textsuperscript{TM}~source codes for all discussed metrics are available at \cite{SourceCode}.

\bibliographystyle{model1-num-names}
\bibliography{SW}{}

\end{document}